\documentclass[proceedings, preprint]{rmaa}

\usepackage{paralist}

\usepackage{psfrag,color}
\usepackage{epstopdf}

\SetYear{2013}
\SetConfTitle{Third AstroRob Workshop}

\title{Late Variability of Flux and Spectra of the Tidal Disruption Flare Sw J1644+57 from \emph {XMM-Newton} data} 

\author{
  A. Gonz\'alez-Rodr\'iguez,\altaffilmark{1}
  A.J. Castro-Tirado,\altaffilmark{1}
  M.A. Guerrero, \altaffilmark{1}
  and A. Castell\'on \altaffilmark{2}}

\altaffiltext{1}{Instituto de Astrof\'isica de Andaluc\'ia, IAA - CSIC. P.O. Box 03004, E-18080 Granada, Spain (agonzalez@iaa.es, ajct@iaa.es, mar@iaa.es).}

\altaffiltext{2}{Facultad de Ciencias, Dpto. de \'Algebra, Geometr\'ia y Topolog\'ia, Universidad de M\'alaga. Campus de Teatinos, s/n M\'alaga, Spain}

\shortauthor{Gonz\'alez-Rodr\'iguez, et.~al.}
\shorttitle{Late Variability of TDF Sw J1644+57}

\listofauthors{Gonz\'alez-Rodr\'iguez, Castro-Tirado, et.~al.}
\indexauthor{Gonz\'alez-Rodr\'iguez, A.}
\indexauthor{Castro-Tirado, A.J.}
\indexauthor{Guerrero, M.A.}
\indexauthor{Castell\'on, A.}

\abstract{We describe the late spectral variability and flux evolution of TDF Sw J1644+57, a Tidal Disruption Flare which left the typical potential trend proportional to $t^{-5/3}$ in 2012, maintaining a quiescent flux until nowadays. Sixteen X-ray observations of ESA satellite \emph{XMM-Newton} have been used in this study, including the one performed on 17$^{\rm th}$July, 2013. A search for optical emission in BOOTES/CASANDRA database has been performed too. Late X-ray fluxes show that the source flux decline does not follow the expected TDF trend at the time of the last \emph{XMM-Newton} observation. Moreover, the spectra fitting parameters, in particular the neutral hydrogen column density, N$_{\rm H}$, and the power-law index, $\Gamma$, indicate that the source darkening has diminished and that the spectral shape has flattened with time. The disruption of the star could have come to an end. Nevertheless, a quiescent X-ray flux continues. Evidence for a quiescent X-ray flux is presented.}

\resumen{Presentamos la variabilidad espectral tard\'ia y la evoluci\'on de flujo del TDF Sw J1644+57, un fen\'omeno transitorio probablemente producido por el desgarramiento de una estrella por fuerzas de marea, el cual en 2012 abandon\'o la tendencia caracter\'istica proporcional a $t^{-5/3}$ para mantener un flujo estable en quietud hasta la actualidad. Para este trabajo se han empleado diecis\'eis observaciones del sat\'elite de la ESA \emph{XMM-Newton}, incluida la \'ultima observaci\'on (17 de Julio de 2013). Asimismo, se ha llevado a cabo una b\'usqueda en la base de datos de BOOTES/CASANDRA a fin de detectar la contrapartida \'optica. Los flujos tard\'ios en rayos X muestran que el decaimiento de flujo de la fuente ya no sigue la tendencia esperada para un TDF. Adem\'as, los par\'ametros del ajuste espectral, en particular la densidad de columna de hidr\'ogeno neutro, N$_{\rm H}$, y el \'indice de la ley de potencias, $\Gamma$, indican que el oscurecimiento de la fuente ha disminuido y que los espectros se han aplanado con el tiempo. El desgarramiento de la estrella podr\'ia haber llegado a su fin. Sin embargo, el flujo de rayos X en quietud contin\'ua. Hay indicacaciones de emisi\'on en quietud.}

\addkeyword{Galaxies: Active}
\addkeyword{ISM: Jets and outflows}
\addkeyword{Stars: Flare}
\addkeyword{X-rays: Galaxies}

\begin{document}
\maketitle

\section{Inroduction}
\label{sec:intro}

Stars in galactic nuclei can be captured or tidally disrupted by a central super massive black hole of $10^{5}-10^{6}\, M_\odot$. The stellar debris that are not captured by the black hole are ejected at high speed, whereas the remainder will be swallowed by the hole.

This phenomenon is known as TDF, Tidal Disruption Flare, and causes a bright flare that usually lasts some years. The radiation can be detected in radio and in X-ray. That is the reason why the TDF Sw J1644+57 has been being monitored by ESA satellite \emph{XMM-Newton}.

The source was discovered the 28$^{th}$ of March of 2011 by \emph{Swift} as a new transient X-ray source, and is thought as the awakening of a quiescent black hole located in the center of the optical position of a host galaxy, at redshift z$=$0.3534 \citep{2011Sci...333..203B, 2011Natur.476..425Z, 2013RMxAC..42...36C}. During 2011 its flux followed a typical TDF potential trend proportional to $t{}^{-\frac{5}{3}}$ \citep{1988Natur.333..523R}.

\section{Observations and Data Reduction}
\label{sec:observations}

There are sixteen X-ray observations of the TDF Sw J1644+57 in the \emph {XMM-Newton} Scientific Archive, XSA, for the coordinates (J2000): RA: 16 44 49.97; Dec: +57 34 59.7.

To make the data reduction and calibration, the Science Analysis System of \emph {XMM-Newton}, SAS, and its Current Calibration Files, CCFs, were employed. X-ray light curves and spectra were extracted once the event lists were generated, cleaned and filtered against high background.

The X-ray Spectral Fitting Package of HEAsoft, XSPEC, has been used to fit the spectra and to calculate fluxes and luminosities. The spectral fittings have provided values for N$_{\rm H}$, the column density of HI in the source direction.

\begin{figure}[t!]
  \includegraphics[width=\columnwidth]{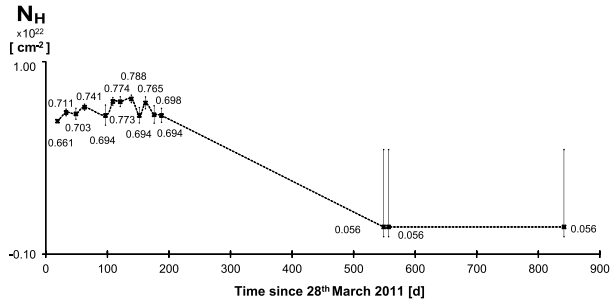}
  \caption{Evolution of the HI column density, N$_{\rm H}$.}
  \label{fig:NH}
\end{figure}

\begin{figure}[b!]
  \includegraphics[width=\columnwidth,height=6.6cm]{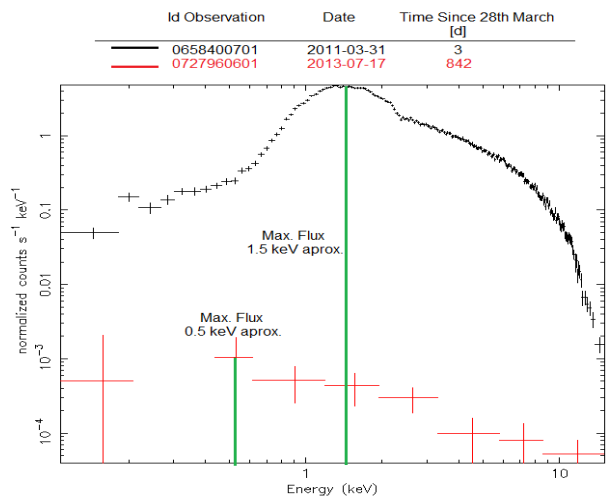}
  \caption{Comparison between the first and the last X-ray spectra.}
  \label{fig:Flattening}
\end{figure}

\section{Results and Discussion}
\label{sec:Results}

\begin{asparaitem}

\item The  N$_{\rm H}$ has decreased as  Figure~\ref{fig:NH} shows, between day 188 and 549, which indicates that most of the clouds of gas and dust around the disrupted star have dissipated. The final number of 0.056$\times10^{22}$cm$^{-2}$ is very close to the galacticN$_{\rm H}$, and implies that the phenomenon is less energetic now.

\item The spectral shapes have flattened with time, as Figure~\ref{fig:Flattening} reveals when we compare the spectrum of the first observation, performed on day 3, with the last one, taken on day 842. Besides, in late observations the maximum flux occurs at a lower energy.

\begin{figure}[!t]
  \includegraphics[width=\columnwidth]{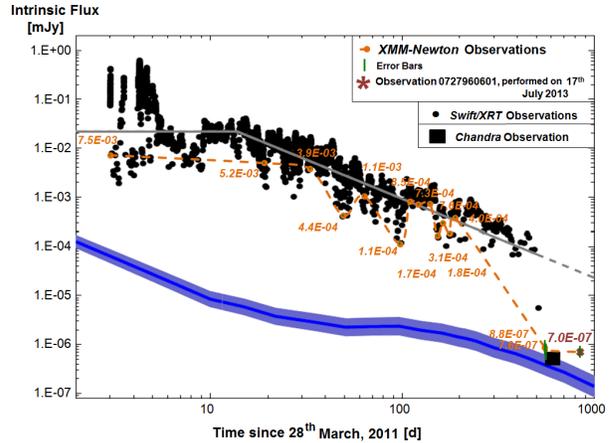}
  \caption{Evolution of the intrinsic flux (0.5$-$8.0 keV). \emph{Swift} observations are drawn as black circles. The most recent \emph{Chandra} observation is shown as a black square. The fluxes obtained in this work are presented as orange circles. In red, at the bottom right, our latest \emph{XMM-Newton} observation. Adapted from \citet{2013ApJ...767..152Z}}
  \label{fig:Intrinsic_Flux}
\end{figure}

\item The intrinsic fluxes between 0.5 and 8.0 keV calculated from the sixteen observations of the \emph{XMM-Newton} Archive, are consistent with previous works about Sw J1644+57. See  Figure~\ref{fig:Intrinsic_Flux}. 

\item Between late 2011 and 2012, the source X-ray flux started to decline, leaving the characteristic potential tendency of a TDF in 2012. Consequently, the black hole has probably used up the star that provided dust and gas clouds, or its remains have escaped from the orbit around the hole \citep{1988Natur.333..523R}.
   	    
\item The object behavior has changed from an X-ray emission compatible with a TDF, to a quiescent emission that remains until nowadays and that is higher than Sagittarius A$^{*}$. We ignore if the accretion already existed before \emph{Swift} triggered.

Thus, the source could be the beginning of an Active Galactic Nucleus, AGN, with the jet faced on, i.e., a mini-blazar. Additional X-ray observations are needed to confirm it.

\item We do find no transient optical emission (brighter than 10$^{\rm th}$ mag) at the CASANDRA all-sky images taken at the different BOOTES robotic astronomical stations world wide in the period 2009-2013 when data is available \citep{2012ASInC...7..313C}.

\end{asparaitem}


\end{document}